%% file: main.tex
\documentclass[twocolumn]{article}
\pdfoutput=1
\usepackage[utf8]{inputenc}
\usepackage[toc, acronym, shortcuts]{glossaries}
\usepackage{csquotes}
\usepackage[english]{babel}
\usepackage[letterpaper]{geometry}
\usepackage{amsmath}
\usepackage{ftnright}
\usepackage{subcaption}
\usepackage{graphicx, float}
\usepackage{refcount}
\usepackage[pdfencoding=auto, unicode, psdextra]{hyperref}
\usepackage{bookmark}
\graphicspath{{\subfix{images/}}}
\usepackage[backend=bibtex]{biblatex}
\usepackage{xr}
\usepackage{subfiles}
\providecommand{\keywords}[1]
{
  \small
  \textbf{\textit{Keywords---}} #1
}

\makeglossaries

\newacronym{tee}{TEE}{Trusted Execution Environment}
\newacronym{pii}{PII}{Personally Identifiable Information}
\newacronym{csp}{CSP}{Cloud Service Provider}
\newacronym{ra}{RA}{remote attestation}
\newacronym{tcb}{TCB}{trusted computing base}
\newacronym{tservice}{TService}{TEE-based service}
\newacronym{tclient}{TClient}{TEE-based client}
\newacronym{atq}{AtQ}{Attest Quote}
\newacronym{atb}{AtB}{Attest Base}
\newacronym{trr}{TrR}{Trusted Request}
\newacronym{utr}{UtR}{Un-trusted Request}
\newacronym{qudd}{QUDD}{Quote User Defined Data}
\newacronym{ahl}{AHL}{Attest Header Line}
\newacronym{ahf}{AHF}{Attest Header Field}
\newacronym{hl}{HL}{Header Line}
\newacronym{hf}{HF}{Header Field}
\newacronym{atr}{AtR}{Attest Request}
\newacronym{ae}{AE}{Authenticated Encryption}
\newacronym{mhttpa}{mHTTPA}{Mutual HTTPA}
\newacronym{mac}{MAC}{Message Authentication Code}
\newacronym{aad}{AAD}{Authenticated Associated Data}
\newacronym{aead}{AEAD}{Authenticated Encryption with Associated Data}
\newacronym{att}{AtT}{Attest Ticket}
\newacronym{atbr}{AtBr}{Attest Binder}
\newacronym{trc}{TrC}{Trusted Cargo}
\newacronym{rtt}{RTT}{round trip time}
\newacronym{tls}{TLS}{Transport Layer Security}
\newacronym{http}{HTTP}{Hypertext Transfer Protocol}
\newacronym{https}{HTTPS}{HTTP Secure}
\newacronym{httpa}{HTTPA}{HTTP Attestable}
\newacronym{ecdhe}{ECDHE}{Ephemeral Elliptic Curve Diffie-Hellman}
\newacronym{(ec)dhe}{(EC)DHE}{Ephemeral (Elliptic Curve) Diffie-Hellman}
\newacronym{hkdf}{HKDF}{HMAC-based Extract-and-Expand Key Derivation Function}
\newacronym{ca}{CA}{certification authority}
\newacronym{pki}{PKI}{public key infrastructure}
\newacronym{svn}{SVN}{security version number}
\newacronym{isv}{ISV}{independent software vendor}
\newacronym{intauth}{IntAuth}{integrity and authenticity}
\newacronym{confintauth}{ConfIntAuth}{confidentiality, integrity and authenticity}
\newacronym{pfs}{PFS}{Perfect Forward Secrecy}
\newacronym{uniqintauth}{UniqIntAuth}{uniqueness, integrity and authenticity}
\newacronym{ak}{AK}{attestation key}
\newacronym{trtls}{TrTLS}{Trusted Transport Layer Security}
\newacronym{hsts}{HSTS}{HTTP Strict Transport Security}
\newacronym[longplural={Roots of Trust}]{rot}{RoT}{Root of Trust}
\newacronym{aths}{AtHS}{Attest Handshake}
\newacronym{atsp}{AtSP}{Attest Secret Provisioning}
\newacronym{ree}{REE}{Rich Execution Environment}
\newacronym{sgx}{Intel\textsuperscript{\textregistered} SGX}{Intel\textsuperscript{\textregistered} Software Guard Extensions}
\newacronym{tdx}{Intel\textsuperscript{\textregistered} TDX}{Intel\textsuperscript{\textregistered} Trust Domain Extensions}
\newacronym{qservice}{QService}{Quoting Service}
\newacronym{uri}{URI}{Uniform Resource Identifier}
\newacronym{mitm}{MITM}{Man-in-the-Middle}
\newacronym{utc}{UTC}{Coordinated Universal Time}
\newacronym{ux}{UX}{user experience}
\newacronym{poc}{PoC}{proof-of-concept}
\newacronym{taas}{TaaS}{Trust-as-a-Service}

\title{HTTPA/2: a Trusted End-to-End Protocol for Web Services}
\author{
Gordon King\footnotemark{}\\
\href{mailto:gordon.king@intel.com}{gordon.king@intel.com}\\
Intel Corporation
\and
Hans Wang\footnotemark[\value{footnote}]{}\\
\href{mailto:hans.wang@intel.com}{hans.wang@intel.com}\\
Intel Corporation
}
\date{}

\begin{document}

\twocolumn[

\begin{@twocolumnfalse}
\maketitle
\hrule
\begin{abstract}
With the advent of cloud computing and the Internet, the commercialized website becomes capable of providing more web services, such as software as a service (SaaS) or function as a service (FaaS), for great user experiences. 
Undoubtedly, web services have been thriving in popularity that will continue growing to serve modern human life. 
As expected, there came the ineluctable need for preserving privacy, enhancing security, and building trust. 
However, HTTPS alone cannot provide a remote attestation for building trust with web services, which remains lacking in trust. 
At the same time, cloud computing is actively adopting the use of TEEs and will demand a web-based protocol for remote attestation with ease of use. 
Here, we propose HTTPA/2 as an upgraded version of HTTP-Attestable (HTTPA) by augmenting existing HTTP to enable end-to-end trusted communication between endpoints at layer 7 (L7). 
HTTPA/2 allows for L7 message protection without relying on TLS. 
In practice, HTTPA/2 is designed to be compatible with the in-network processing of the modern cloud infrastructure, including L7 gateway, L7 load balancer, caching, etc.  
We envision that \acs{httpa}/2 will further enable trustworthy web services and trustworthy AI applications in the future, accelerating the transformation of the web-based digital world to be more trustworthy.
\medskip \\
\begingroup
\centering
\endgroup
\smallskip \\
\hrule
\bigskip
\end{abstract}
\end{@twocolumnfalse}
]

\def\thefootnote{*}\footnotetext[\value{footnote}]{Both authors contributed equally to this work.}

\section{Introduction}\label{intro}
\subfile{tex/intro}

\section{Technical Preliminaries}\label{preliminaries}
\subfile{tex/prelim}
\section{Protocol Transactions}\label{protointeraction}

\subfile{tex/interaction}

\section{Security Considerations}\label{secconsider}
\subfile{tex/security}

\section{Conclusion}\label{conclusion}
\subfile{tex/conclusion}

\section{Future Work}\label{futurework}
\subfile{tex/futurework}

\section{Acknowledgements}\label{acknowledges}
We would like to acknowledge the support from the \acs{httpa} workgroup members, including our partners and reviewers. 
We thank their valuable feedback and suggestions.

\section{Notices \& Disclaimers}\label{disclaimers}
No product or component can be absolutely secure.

\clearpage
\printglossary
\printglossary[type=\acronymtype]

\clearpage
\printbibliography
\end{document}

%% file: tex/intro.tex
We received positive feedback and inquiries on the previous version of \acs{httpa}~\cite{orighttpa} (\acs{httpa}/1).
As a result, we present a major revision of the \acs{httpa} protocol (\acs{httpa}/2) to protect sensitive data in \acs{httpa} transactions from cyber attacks.
Comparatively, the previous work ~\cite{orighttpa} is mainly focused on how to include \ac{ra} and secret provisioning to the \acs{http} protocol with \ac{tls} protection across the Internet, which is great, but it comes at a price.
In contrast, \acs{httpa}/2 is not necessary to rely on the \ac{tls} protocol, such as \acs{tls} 1.3~\cite{rfc8446}, for secure communication over the Internet.
The design of \acs{httpa}/2 follows the SIGMA model~\cite{sigma_model} to establish a trusted (attested) and secure communication context between endpoints at layer 7 (L7) of the OSI model.
Different from connection-based protocols, \acs{httpa}/2 is transaction-based in which the \acs{tee} is considered to be a new type of requested resource over the Internet.
In addition to protecting sensitive data transmitted to \acp{tservice}, \acs{httpa}/2 can potentially be used to optimize the end-to-end performance of Internet or cloud backend traffic, thus saving energy and reducing the operational costs of \acp{csp}.

\acs{http} is a predominant Layer 7 protocol for website traffic on the Internet.
The \acs{httpa}/1~\cite{orighttpa} defines an \acs{http} extension to handle requests for remote attestation, secret provisioning and private data transmission, so Internet visitors can access a wide variety of services running in \acp{tee}~\cite{jauernig2020trusted} to handle their requests with strong assurances.
In this way, visitors' \ac{pii} and their private data are better protected when being transmitted from a client endpoint to a trusted service endpoint inside the \acs{tee}.
The \acs{httpa}/1 supports mutual attestation if both client and service endpoints run inside the \acs{tee}.
Although \acs{httpa}/1 helps build trust between L7 endpoints with data-level protection, \acs{httpa}/1 needs \acs{tls} to defend against attacks over the Internet, e.g., replay attacks and downgrade attacks.
Note that \acs{tls} cannot guarantee end-to-end security for the \acs{https} message exchange~\cite{httpsig} when the \acs{tservice} is hosted behind a \acs{tls} termination gateway or inspection appliance (a.k.a. middle boxes).
Despite the fact that \acs{tls} provides \ac{confintauth} to ensure secure message exchange for the \acs{httpa}/1 protocol, it is not a complete end-to-end security solution to serve web services at L7. 
For example, \acs{tls} termination on the middleboxes makes it highly vulnerable to cyber-attacks.  
Both \acs{httpa}/1 and \acs{tls} need to generate key material through key exchange and derivation processes. 
This requires additional round trips at L5 and increases network latency.
Thus, there is room to further optimize the network performance and reduce communication complexity by avoiding the repetition of key negotiations.
Due to the limitation of \acs{tls} mentioned above, a version of \acs{httpa} with message-level security protection is a natural candidate to address the issues mentioned above at once.\\
This paper proposes an upgraded protocol, \acs{httpa}/2, which makes it possible to secure \acs{httpa} transactions even with no underlying presence of \acs{tls}.
\acs{httpa}/2 is designed to improve the processes of key exchange, \ac{ra} and secret provisioning in \acs{httpa}/1. 
It also enables end-to-end secure and trustworthy request/response transactions at L7, which is cryptographically bound to an attestable service base that can be trusted by Internet visitors regardless of the presence of untrusted \acs{tls} termination in between.\\
The rest of the paper is organized as follows. Section~\ref{preliminaries} provides necessary preliminaries. Section~\ref{protointeraction} elaborates on the protocol transactions. Section~\ref{secconsider} talks about security considerations. Section~\ref{conclusion} concludes the whole paper.

%% file: tex/prelim.tex
This section provides preliminaries related to the construction of \acs{httpa}/2. We first introduce using a \nameref{attesttee} in a web service setting. Then we describe several important primitives to be used for constructing the \acs{httpa} protocol described in section~\ref{protointeraction}.

\subsection{Trusted Execution Environment (TEE)}\label{attesttee}
In a \acs{tee}, trustworthy code is executed on data with CPU-level isolation and memory encryption inaccessible to anyone even those with system-level privileges. 
The computation inside the \acs{tee} is protected with confidentiality and integrity.
A \acs{tee} is more trustworthy than a \ac{ree}~\cite{ree} where codes are executed on data without isolation.
Although most web services are deployed in \acs{ree}, it is an emerging trend to deploy web services in a \acs{tee} for better security. 
Upon implementation, a service initialized and run inside the \acs{tee} is known as a Trusted Service, called a \ac{tservice}.
\ac{tservice} uses \acs{tee} to effectively reduce its \ac{tcb}, which is the minimal totality of hardware, software, or firmware that must be trusted for security requirements, thus reducing the attack surface to as minimal as possible.
 \\
For some \acs{tee}, such as \ac{sgx} or \ac{tdx}, it can provide evidence (or we call the evidence ``\ac{atq}'' in the section~\ref{attestquote}) reflecting configuration and identities~\cite{flexible_ra} to the remote relying party.  
After successfully verifying the evidence and being convinced of the result, both parties finish the \ac{ra} process~\cite{ra_review_2021}~\cite{remoteattestsrv}.
With the \ac{ra} completed successfully, \acp{tservice} can be shown more trustworthy to its relying party.\\
Not all \acsp{tee} are attestable, and the \acs{httpa} is only applicable to attestable \acs{tee} which can generate such evidence for the purpose of \ac{ra}.

\subsection{Attest Quote (AtQ)}\label{attestquote}
\acs{atq} is an opaque data structure signed by a \ac{qservice} with an \ac{ak}, and it can be called a quote~\cite{explore_ra_tee} or evidence, which is used to establish trustworthiness through identities. Because of this, the quote encapsulates code identity, \acs{isv} identity, \acs{tee} identity, and various security attributes, e.g., \ac{svn} of a \ac{tee}~\cite{innocpuattest}, associated with a \acs{tservice} instance. A relying party can examine the quote to determine whether the \acs{tservice} is trustworthy or not via verification infrastructure.\\
 The quote is not a secret, and it must ensure its \ac{uniqintauth}. 
 The quote generation involves cryptographically measuring the instantiated \ac{tcb}, signing the measurements with an \acs{ak}, including a nonce.\\
 The \acs{atq} accommodates a piece of user-defined information, called \ac{qudd}. The \acs{qudd} can provide extra identities specific to a \acs{tservice}. Therefore, the \acs{atq} can in turn help protect the integrity of \acp{ahl} during the handshake phase which we will discuss in section~\ref{sec:raalloc}.\\
 It's worth noting that not all quotes are the same, especially if they are structured by different vendors, so a label of quote type should be attached along with \acs{atq}.

\subsection{Attest Base (AtB)}\label{attestbase}
\acs{atb} is the totality of computing resources serving client request handling, including hardware, firmware, software, and access controls to work together to deliver trustworthy service quality with enforced security/privacy policy.
It can also be considered as a group of collaborating \acs{tservice} instances running in their own attestable \acsp{tee} respectively, which are capable of proving the integrity of the execution state.\\
In this paper, the computing resources of \acs{tservice} are offered by \acs{atb} to be accessible to the client through a series of trusted transactions.
How to attest those trustworthy services is determined by the specified policy in the handshake phase.
We suggest using a single \acs{tservice} instance for each \acs{atb} to reduce the complexity and attack surface as much as possible.\\
The \acs{atb}, serving for a particular service tied to a \ac{uri}, should be directly or indirectly attested by a client through \acs{httpa}/2 protocol.\\
In the case of a \acs{atb} formed by multiple \acsp{tservice} instances, an upfront \acsp{tservice} instance takes responsibility for performing local attestation on the rest of \acsp{tservice} instances to establish trustworthy relationships with them.
After that, the upfront \acsp{tservice} can selectively collect their quotes for client-side verification during the \acs{httpa}/2 handshake phase.

\subsection{Three Types of Request}
There are three types of request defined by the \acs{httpa}/2 protocol, including \ac{utr} \ac{atr}, and \ac{trr}. 
\acs{utr} is used in \acs{http} transactions; \acs{atr} is used in both transactions of \ac{aths} and \ac{atsp}; \acs{trr} is used in trusted transaction.
For convenience we refer to the \acs{atr} and \acs{trr} as ``\acs{httpa}/2 request''.\\
Regarding \acs{http} method, we propose a new \acs{http} method, called ``ATTEST'', to perform the transactions of \ac{aths} and \ac{atsp}. 
The \acs{http} request using ATTEST method is called \acs{atr}.
Regarding \acs{http} header fields, we propose to augment them with additional ones called \acp{ahf} prefixed with string ``Attest-''. 
Without \acp{ahf}, it must be a \ac{utr} in terms of \acs{httpa}/2.\\

The \acp{ahf} are dedicated to \acs{httpa} traffic. 
For example, they can be used to authenticate the identity of \acs{httpa}/2 transactions, indicate which \acs{atb} to request, convey confidential meta-data (see section~\ref{trustcargo}), provision secrets, present ticket (see section~\ref{attestticket}), etc.\\
Last one is \ac{ahl}, it consists of \ac{ahf} and its values in a standard form~\cite{rfc8941}. 
We use it to signify a single piece of annotated data associated with the current \acs{httpa}/2 request.

\subsubsection{Un-trusted Request (UtR)}\label{untrustreq}
The \acs{utr} is simply an ordinary type of \acs{http} request, which does not use ATTEST method nor does it contain any \acsp{ahl}.\\
Before a \acs{utr} reaches a \acs{tservice}, the \acs{utr} can be easily eavesdropped on or tampered with along the communication path. 
Even protected by \acs{tls}, it is still possible to be attacked when crossing any application gateway or L7 firewall since those intervening middle-boxes are un-trusted and will terminate TLS connections hop by hop~\cite{httpsig}. 
Therefore, there is no guarantee of \ac{confintauth}. 
That's why the \acs{tservice} cannot treat the request as trustworthy, but it is still possible for \acs{tservice} to handle \acs{utr} if allowed by the service-side policy.
Thus, we don't suggest \acs{tservice} to handle any one of them for the sake of security.

\subsubsection{Attest Request (AtR)}\label{attestreq}
The \acs{atr} is an \acs{http} request equipped with both ATTEST method and \acsp{ahl} for \ac{aths} and \ac{atsp}. 
If any \acs{atr} was not successfully handled by corresponding \acs{tservice}, subsequent \acs{trr}, described in the section~\ref{trustreq}, will no longer be acceptable to this \acs{tservice}.
We describe the major difference between an \acs{atr} used in \acs{aths} and \acs{atsp} respectively as follows:
\paragraph{\ac{aths}}
The \acs{atr} used in \ac{aths} is designed to request all necessary resources for handling both types of \acs{atr} used in \ac{atsp} and \ac{trr}. 
For example, one of the most important resources is \acs{atb} (see section~\ref{attestbase}), which may be scheduled or allocated by a server-side resource arbiter. Typically (but not always), an upfront \acs{tservice} can directly designate itself as the \acs{atb} for this client. 
For the complete explanation in detail, see section~\ref{sec:raalloc}.\\
In addition, this \acs{atr} should not be used to carry any confidential information, because the key material cannot be derived at this moment. 
\acs{tservice} can encrypt sensitive data in the response message since it has already received the required key share from the client and be able to derive the key material for encryption. 

\paragraph{\ac{atsp}}
The \acs{atr} of \ac{atsp} is optional and may not be present in \acs{httpa}/2 traffic flow since in some cases the \acs{tservice} does not need any \acs{atb}-wide secrets provided by the client to work.
In the common case, \acs{tservice} needs secret provisioning to configure its working environment, such as connecting to databases, setup signing keys, and certificates, etc. 
This \acs{atr} must be issued after all \acs{tee} resources have been allocated through the \ac{aths} transaction described above.\\
It's worth noting that this request is not required to be issued before any \ac{trr}~\ref{trustreq}.
With such flexibility, \acs{tservice} can get extra information or do some operations beforehand through preceding \acsp{trr}.\\
Importantly, this \acs{atr} is responsible to provision \acs{atb}-wide secrets to \acs{atb}, such as key credentials, tokens, passwords, etc. 
Those secrets will be wrapped by an encryption key derived from the key exchange in the \acs{aths} phase~\ref{sec:raalloc}. 
Furthermore, the \acs{tservice} must ensure that those provisioned secrets will be eliminated after use, \acs{atb} get released, or any failure occurred in processing \acs{atr}. 
For the complete explanation in detail, see section~\ref{ssec:secprov}.

\paragraph{}
The two kinds of \acs{atr} introduced above are the core of \acs{httpa}/2 protocol. 
Both of them can be treated as GET request messages to save a \acs{rtt}, and they can also be used to transmit the protected sensitive data to both sides, except for the \acs{atr} of \ac{aths} due to the key exchange not yet completed as noted earlier.

\subsubsection{Trusted Request (TrR)}\label{trustreq}
The \acs{trr} can be issued right after successful \acs{aths} (see section~\ref{attestreq}) where an \acs{atb} is allocated.
Although, \acs{trr} does not use ATTEST method, it should contain \acsp{ahl} to indicate that it is a \acs{trr} not a \acs{utr}. 
In other words, the \acs{trr} is nothing but an ordinary \acs{http} request with some \acsp{ahl}. 
Within those \acsp{ahl}, one of them must be \acs{atb} ID to determine which \acs{atb} is targeted in addition to the specified \acs{uri}. 
With that, the \acs{trr} can be dispatched to proper \acs{tservice} to handle this request.\\
In essence, the \acs{trr} is designed to interact with a \acs{tservice} for sensitive data processing. 
The \acs{httpa}/2 must ensure \acs{confintauth} of a set of selected data, which may be distributed within the message body or even in the header or request line, for end-to-end protection.
It turns out that not all message bytes would be protected under \acs{httpa}/2 like \acs{tls} does.
As a result, the \acs{httpa}/2 may not be suitable for certain scenarios, e.g., simply want to encrypt all traffic bytes hop by hop. 
However, in most cases, the \acs{httpa}/2 can offer a number of obvious benefits without \acs{tls}. 
For example, users do not need to worry about data leakage due to \acs{tls} termination, and they can save the resources required by the \acs{tls} connections. 
In some special cases, the \acs{httpa}/2 can be combined with \acs{tls} to provide stronger protection but the performance overhead can be significant (see section~\ref{trustls}).\\
There are many potential ways to optimize Internet service infrastructure/platform by means of adopting \acs{httpa}/2 since the insensitive part of \acs{httpa}/2 messages can be used as useful hints to improve the efficiency of message caching or routing/dispatching, risk monitoring, malicious message detection and so on, helping protect sensitive data in motion, as well as in processing by the client-chosen \acsp{tservice}.

\subsection{Attest Ticket (AtT)}\label{attestticket}
\acs{att} is a type of \acs{ahl} used to ensure the \ac{intauth} of \acsp{ahl} and freshness by applying \acs{aad} to each \acs{httpa}/2 request, except for the \acs{atr} of \acs{aths} which is the initiating request for the handshake.
\acs{att} is required to be unique for single use so as to mitigate the replay attack as it can be ensured by \acs{aad} in practice.
Moreover, the \acs{att} should be appended at the very end of the request body as the last trailer~\cite{rfc7230} because there might be \acs{trc} or other trailers which need to be protected by the \acs{att} as well.\\
Regarding the \acs{atr} of \ac{aths}, there is no protection from \acs{att}, because there are no derived keys available to use at such an early stage.
In order to protect the \acs{atr} of \ac{aths}, we can use either client-side quote or pre-configured signing key methods to ensure the \ac{intauth} instead of \acs{att}. 
Typically, there are four situations to consider: \acs{mhttpa}, client with \acs{ca}-signed certificate, client with self-signed certificate, and nothing to provide.
\paragraph{\ac{mhttpa}} 
With mutual \acs{httpa} being used, the client must be running on a \acs{tee} as \ac{tclient}, which is capable of generating a client-side \acs{atq}.
The \acs{atq} can be used to ensure the \acs{intauth} of \acs{atr} by means of including the digest of \acsp{ahl} into its \ac{qudd}, and the server-side should have a proper trusted attestation authority to verify it.\\
This is the recommended approach to build mutual trust between \acs{tclient} and \acs{tservice}, but the client-side usually lacks of \acs{tee} feature support.

\paragraph{Client with \acs{ca}-signed Certificate} 
In this case, the client signs the \acsp{ahl} of \ac{atr} along with a trusted certificate, and \acs{tservice} should be able to verify the signature with respective \acs{ca} certificate chain.
This way helps \acs{tservice} to identify the user identity.\\
In addition, the \ac{mhttpa} can be enabled at the same time to make it more secure and trustworthy on both sides if possible.

\paragraph{Client with self-signed Certificate} 
In this situation, the client should sign the \acsp{ahl} of \acs{atr} using temporary signing key, and \acs{tservice} should verify the signature using its self-signed certificate enclosed in the same \acs{ahl}. 
This approach is not safe since the \acs{tservice} may receive compromised \acs{atr}.
    
\paragraph{Nothing to provide} 
There is no way to protect the integrity of \acsp{ahl} under these circumstances. 
We recommend at least using the temporary generated signing key with the corresponding self-signed certificate. 
It's worth noting that even if the \acsp{ahl} of \acs{att} is compromised, the client is able to detect the problem by checking the received \acs{atq} of the \acs{tservice} as the \acs{qudd} embedded in \acs{atq} will ensure the integrity of \acsp{ahl} in its request and response messages altogether.\\

%
It is difficult for the client to simply use the self-signed certificate to prove its identity, let alone in the case of nothing to provide.
Again we recommend the client to combine \acs{mhttpa} and \acs{ca} signed certificate approaches together to establish strong trustworthy relationship between \acs{tclient} and \acs{tservice} if the server also wants to identify the client's identity at the initiating request, \acs{atr}. 
If it is not the case, the client must detect whether any unexpected changes occurred in the \acsp{ahl} of \acs{atr} as an additional critical step to defend against \ac{mitm} and downgrade attacks.

\subsection{Attest Binder (AtBr)}\label{attestbinder}
\acs{atbr} is a type of \acs{ahl} used to ensure the binding between \acs{httpa}/2 request and the corresponding response. 
The \acs{ahl} of \acs{atbr} should be added into the response message as the last trailer~\cite{rfc7230}.
The \acs{atbr} typically holds the \ac{mac} to protect two components: all \acsp{ahl} of the current response, and the value of \acs{att} in its corresponding request.
We can choose other cryptographic algorithms for encryption and message authentication, e.g., \ac{aad}, \ac{aead}~\cite{rfc5116} to ensure the \acs{intauth} of the \acs{atbr}.\\
The \acs{atbr} should present in all \acs{httpa}/2 response messages, except for the response of \acs{atr} in the \acs{aths} phase~\ref{sec:raalloc}. 
The reason is the quote of \acs{tservice} can achieve the same purpose without the help of \acs{atbr}.

\subsection{Trusted Cargo (TrC)}\label{trustcargo}
The \acs{trc} can appear in both of \acs{httpa}/2 request and response messages, except for the \acs{atr} of \acs{aths}.
The \acs{trc} serves as a vehicle to carry confidential information which needs to be protected by authenticated encryption.\\
\acs{trc} can be used to protect some sensitive metadata such as data type, location, size, and key index to tell the places in which the ciphertext or signed plaintext is located in the message body.
The key index indicates which key should be used to decrypt those encrypted messages or verify the message's integrity.
Potentially, there is much useful metadata that can be included in \acs{trc}, but we should keep it minimum as size limits might be enforced by intermediaries in the network path.\\
The way to structure the metadata and how to parse it is not defined by this paper. 
We leave it to the future extensions of \acs{httpa}/2 or it can be customized by application.\\
Finally, the \acs{trc} should be put in a trailer~\cite{rfc7230} since its variable length affects the position information it contains. 
Again, the value of \acs{trc} must be protected from eavesdropping on or manipulating by the means of \acs{ae}.

\subsection{Trusted Transport Layer Security (TrTLS)}\label{trustls}
\acs{httpa}/2 protects the selected parts of an HTTP message.
If users want to protect the entire HTTP message---every bit of the message, \acs{tls} can leverage \acs{httpa}/2 to establish a secure connection at L5 between the client and its adjacent middlebox, which we call \acs{trtls}. 
The \acs{trtls} makes use of \acs{aths} to transmit HELLO messages of \acs{tls}~\cite{rfc8446} to the client and \acs{tservice} respectively for handshake.
This way can make the initial handshake of the secure transport layer protocol trustworthy. 
We consider three cases of endpoints as follows:
\paragraph{\acs{tclient}}
In the case of \acs{tclient} endpoint, the \acs{tclient} leverages its \acs{qudd} to ensure the \acs{intauth} of the client hello message to establish \acs{tls} connection.
The server-side verifier helps verify the \acs{qudd}. 
If the attestation is successful, the trusted endpoint of \acs{tclient} will be established as \acs{trtls}.
Note that the \acs{trtls} module should be co-located in the same \acs{tee} with \acs{tclient}.

\paragraph{Frontend \acs{tservice}}
The frontend \acs{tservice} is defined as a \acs{tservice} which can communicate with the client without any L7 middleboxes in between. 
In other words, the communication between the \acs{tservice} and the client has no \acs{tls} terminators.
This implies that \acs{tservice} can establish a secure transport layer connection directly with its client. 
Thus, the \acs{intauth} of the server-side HELLO message can be fully protected by the \acs{qudd} of \acs{tservice} in a similar way described in the above case of \acs{tclient} but in the reverse direction.
  
\paragraph{Backend \acs{tservice}}
In the case of a backend \acs{tservice} endpoint, the connection of the secure transport layer will be terminated by at least one middle-box e.g., application gateway, reverse proxy, or L7 load balancer. 
Although \acs{tservice} has no direct connection with its client, the trusted connection of \acs{tls} between the client and the first middlebox can be established by checking the results of \acs{ra} from the backend \acs{tservice}.
The middlebox needs to consider the mapping of the request and the response with the backend \acs{tservice} to be correct in order to decide whether to use the results of \acs{ra} to build a trusted connection or not. 
Admittedly, this is the least trustworthy configuration in terms of full traffic encryption when it is compared with the two cases mentioned above because the attack surface includes those vulnerable middleboxes in the \acs{trtls} connection.
\paragraph{}
After the initial message is exchanged through \acs{aths}, the encrypted channel can be established under the \acs{httpa}/2 at L5, so the following traffic will be encrypted and inherit the attested assurances of the \acs{tee} from \acs{httpa}/2.
In the case of the ordinary \acs{tls} connection prior to \acs{httpa}/2, the \acs{trtls} mechanism can disconnect the already built \acs{tls} connection and then re-establish a trustworthy \acs{tls} connection seamlessly.  
We simply present the high-level concept of \acs{trtls} in this paper and may discuss more details in another paper.

%% file: tex/interaction.tex
In this section, we provide detailed definitions of all \acs{httpa}/2 transactions.

\subsection{Preflight Check Phase}\label{preflightcheck}
The preflight request gives the Web service a chance to see what the actual \acs{atr} looks like before it is made, so the service can decide whether it is acceptable or not.
In addition, the client endpoint performs the preflight check as a security measure to ensure that the visiting service can understand the ATTEST method, \acsp{ahf}, and its implied security assurance.\\
To start with \acs{httpa}/2, a preflight request could be issued by a client as optional to check whether the Web service, specified by \acs{uri} in the request line, is \acs{tee}-aware and prepare for \acs{aths}.
In the case that the client is a Web browser, the preflight request can be automatically issued when the \acs{atr} qualifies as ``to be preflighted''.\\
The reason we need the preflight transaction is that it is a lightweight \acs{http} OPTIONS~\cite{httpsemantics} request, which will not consume a lot of computing resources to handle, compared to the \acs{atr}.
Caching the preflight result can avoid the re-check operation during a specified time window.\\

Passing this check does not guarantee that the \acs{atr} can be successfully handled by this service. 
For example, the \acs{tservice} may run out of resources, or the client's cipher suites are not supported, and so on.\\
The preflight is an idempotent operation, e.g., there is no additional effect if it is called more than once with the same input parameters.
The client can also use the preflight to detect the capabilities of \acs{atb}~\ref{attestbase}, without implying any real actions.

\begin{figure*}[htp]
    \centering
    \includegraphics[width=0.8\textwidth]{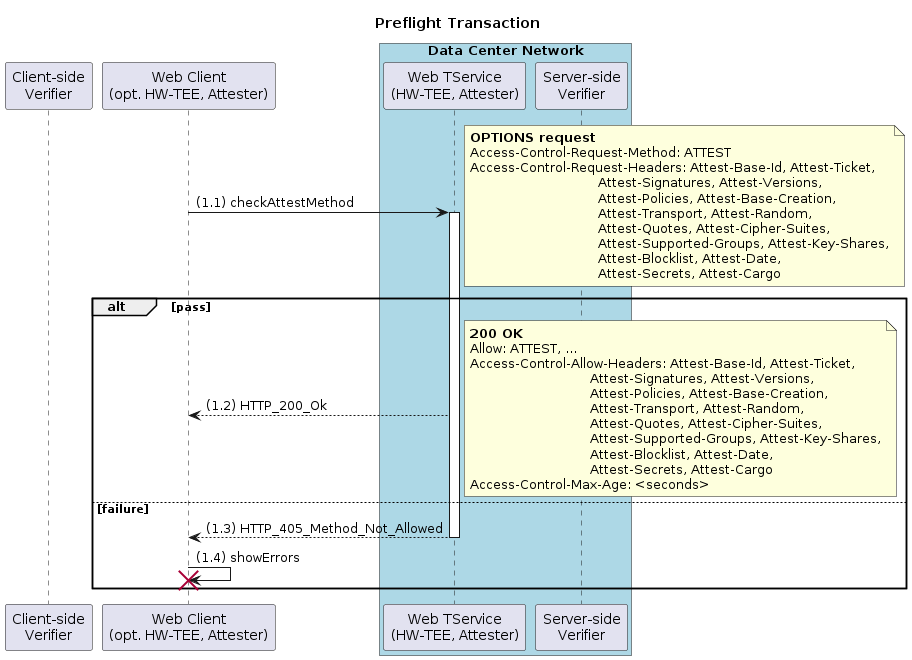}
    \caption{Preflight transaction}
    \label{fig:preflight}
\end{figure*}

As shown in Figure.\ref{fig:preflight}, an OPTIONS request should be honored by an \acs{httpa}/2 compliant \acs{tservice}.
In the preflight transaction, it has standard \acsp{hf} to specify the method and \acsp{ahl} which will be sent out later to the same \acs{tservice} if they are acceptable. Those \acsp{hf} are described respectively as follows:

\begin{enumerate}
    \item \acsp{hf} in request message
    \begin{enumerate}
        \item Access-Control-Request-Method\\
            This \acs{hf} carries a list of methods indicating that ATTEST method will be used in the next request if the service can support it.
        \item Access-Control-Request-Headers\\
            This \acs{hf} carries a list of field names indicating that the \acsp{ahf} will be included in the next request if the service can support it.
    \end{enumerate}
    \item \acsp{hf} in response message
    \begin{enumerate}
        \item Allow\\
            This \acs{hf} carries a list of supported methods by the visiting service.
            It must contain the ATTEST method for the client to proceed with \acs{atr}; otherwise, the \acs{atr} is not acceptable by this service and will be denied if received it. 
        \item Access-Control-Allow-Headers\\
            This \acs{hf} carries a list of allowed \acsp{ahf}. The client needs to check that all of the requested \acsp{ahf} should be contained in this resulting field.
        \item Access-Control-Max-Age\\
            This \acs{hf} indicates how long the preflight check results can be cached.
    \end{enumerate}    
\end{enumerate}

\subsection{Attest Handshake (AtHS) Phase}\label{sec:raalloc}

\begin{figure*}[htp]
    \centering
    \includegraphics[width=0.8\textwidth]{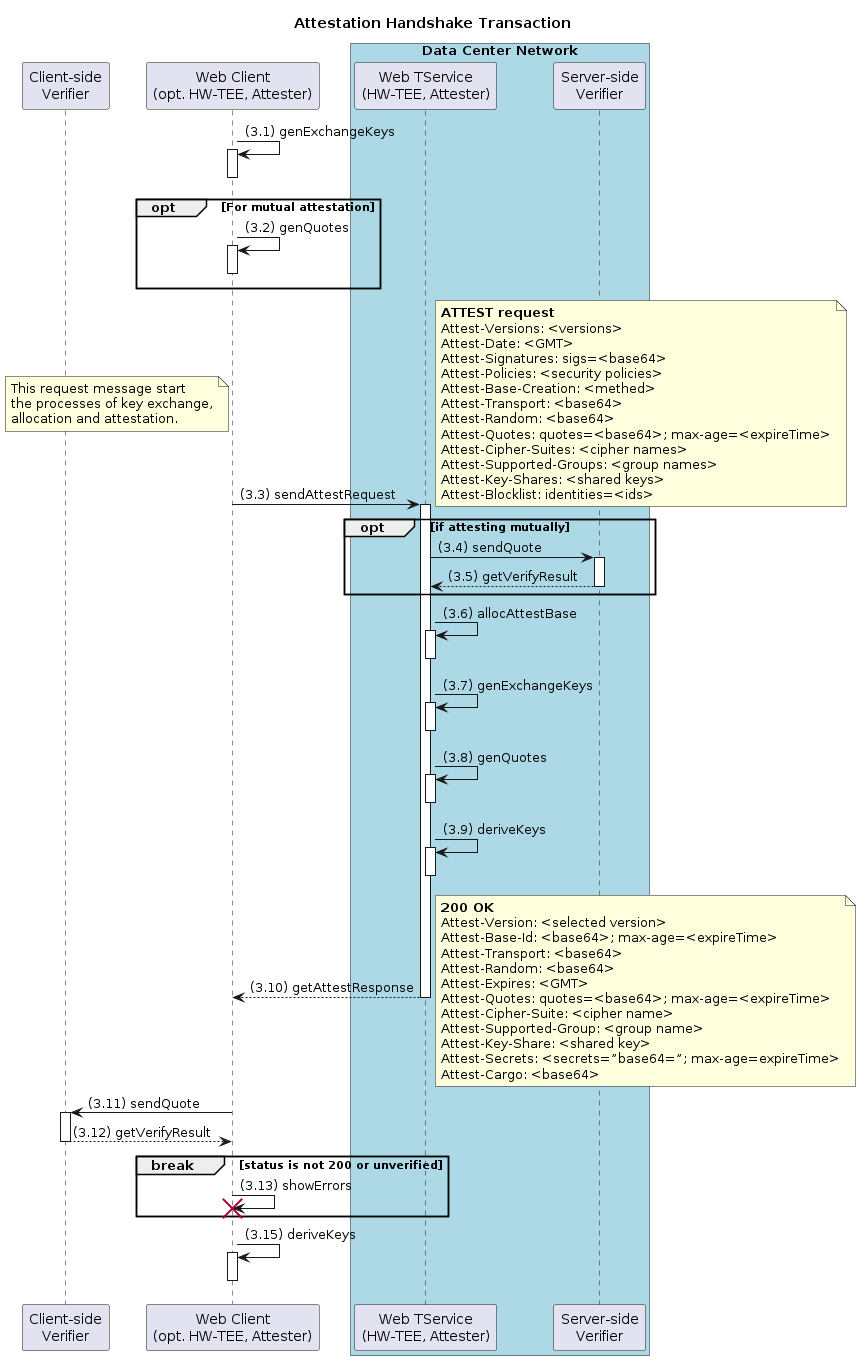}
    \caption{Attest handshake (\acs{aths}) transaction}
    \label{fig:raalloc}
\end{figure*}

The \ac{aths} phase contains a core transaction of \acs{httpa}/2.
In a single round trip time (one \acs{rtt}), the \acs{atr} and its response accomplishes three major tasks, including key exchange, \acs{atb} allocation and \acs{atq} exchange, as shown in Figure.\ref{fig:raalloc}:
\begin{enumerate}
    \item Key Exchange\\
        It is necessary to complete the key exchange process before any sensitive information can be transmitted between the client and \acs{tservice}. 
        The exact steps within this will vary depending upon the kind of key exchange algorithm used and the cipher suites supported by both sides.\\
        In \acs{httpa}/2, the key exchange process follows \acs{tls} 1.3~\cite{rfc8446} and recommends a set of key exchange methods to meet evolving needs for stronger security.\\
        Insecure cipher suites have been excluded, and all public-key-based key exchange mechanisms now provide \ac{pfs}, e.g., \ac{ecdhe}. 
        Note that it is mandatory that the fresh ephemeral keys are generated and used, and destroyed afterward~\cite{rfc8422} inside the \acs{tee} of \acs{tservice}.\\
        When the key exchange is completed, we recommend using \ac{hkdf}~\cite{rfc5869} as an underlying primitive for key derivation.\\
        We describe the key negotiation between the client and the \acs{tservice} in terms of \acsp{ahf} set in request and response respectively as follows:
        \begin{enumerate}
            \item \acsp{ahf} in request message (or \acs{atr}):
            \begin{enumerate}
                \item Attest-Cipher-Suites\\
                It is a list of cipher suites that indicates the \acs{aead} algorithm/\acs{hkdf} hash pairs supported by the client.
            
                \item Attest-Supported-Groups\\
                It is a list of named groups~\cite{rfc7748} that indicates the \acs{(ec)dhe} groups supported by the client for key exchange, ordered from most preferred to least preferred. 
                The \acs{ahl} of Attest-Key-Shares contains corresponding \acs{(ec)dhe} key shares e.g., pubkeys for some or all of these groups.
                
                \item Attest-Key-Shares\\
                Its value contains a list of the client's cryptographic parameters for possible supported groups indicated in the \acs{ahl} of Attest-Supported-Groups for negotiation. 
                We can refer to the corresponding data structure described in \acs{tls} 1.3~\cite{rfc8446}. 
                It is a time-consuming operation to generate those parameters (see 3.1 in Figure.\ref{fig:raalloc})
            
                \item Attest-Random\\
                It is 32 bytes of a random nonce, which is used to derive the master secret and other key materials by \acs{tservice}. 
                The purpose of the random nonce is to bind the master secret and the keys to this particular handshake. 
                This way mitigates the replay attack to the handshake as long as each peer properly generates this random nonce.
            \end{enumerate}
            
            \item \acsp{ahf} in response message
            \begin{enumerate}
                \item Attest-Cipher-Suite\\
                It indicates the selected cipher suites, i.e. a symmetric cipher/\acs{hkdf} hash pair for \acs{httpa}/2 message protection.
                
                \item Attest-Supported-Group\\
                It indicates the selected named group to exchange \acs{ecdhe} key share generated by the \acs{tservice}.
                
                \item Attest-Key-Share\\
                Its value contains the \acs{tservice}'s cryptographic parameters accordingly (see 3.7 in Figure.\ref{fig:raalloc}). 
                
                \item Attest-Random\\
                It takes the same mechanism as the Attest-Random in the request.
                Instead, it is used by the client to derive the master secret and other key materials.
            \end{enumerate}
        \end{enumerate}
        This handshake establishes one or more input secrets combined to create the actual keying materials. 
        The key derivation process (see 3.9, 3.15 in Figure.\ref{fig:raalloc}), which makes use of \acs{hkdf}, incorporates both the input secrets and the \acsp{ahl} of handshake.
        Note that anyone can observe this handshake process if it is not protected by the byte-to-byte encryption at L5, but it is safe since the secrets of the key exchange process will never be sent over the wire.

    \item \acs{atb} Allocation\\
        This task takes care of resource allocation. The upfront \acs{tservice} needs to prepare essential resources before assigning an unique \acs{atb} identifier to the \acs{atb}, which is used by the client to ask \acs{tservice} to process its sensitive data on this \acs{atb} (see 3.6 in Figure.\ref{fig:raalloc}).
        \begin{enumerate}
            \item \acsp{ahf} in request message or \acs{atr}:
            \begin{enumerate}
                \item Attest-Policies\\
                It can contain various types of security policies, which can be selectively supported by this \acs{atb} of \acs{tservice}. 
                There are two aspects to consider as follows:
                \begin{description}
                    \item[Instances attestation] \hfill \\
                    \emph{direct}: all instances should be verified by the client.\\
                    \emph{indirect}: only the contact instance should be verified by the client remotely.
                    
                    \item[Un-trusted requests] \hfill \\
                    \emph{allowUntrustedReq}: it allows \acs{utr} to be handled by the \acs{tservice} on this \acs{atb} (disabled by default).
                \end{description}
                \item Attest-Base-Creation\\
                It specifies a method used for the creation of \acs{atb}. 
                There might be several options available to select:
                \begin{description}
                    \item[new] \hfill \\
                    It means that the \acs{atb} should be newly created for the client to use. If the contact \acs{tservice} is a new one, then it can be assigned to this client immediately.
                    
                    \item[reuse] \hfill \\
                    This option allows reusable \acs{atb} to be used by this client, but the \acs{atb} should ensure that all traces associated with the previous client are erased.\\
                    So far, there is no such \acs{tee}, which can achieve this security feature strictly, and we cannot fully rely on software to emulate it. As a result, the client should evaluate the risks before specifying this option.
                    
                    \item[shared] \hfill \\
                    A shareable \acs{atb} can be allocated to this client. The client does not care whether it is a clean base or not. Use it with caution.
                \end{description}
                \item Attest-Blocklist\\
                It indicates a list of blocked identities and other types of identifiers, which allows \acs{tservice} to filter out unqualified \acs{atb} beforehand. 
                This feature is used to optimize the performance of \acs{atb} allocation, as it is quite expensive and inefficient to rely only on the client to collect a set of \acs{tservice} instances by using the trial and error method.
            \end{enumerate}
            
            \item \acsp{ahf} in response message:
            \begin{enumerate}
                \item Attest-Base-ID\\
                This identifier signifies the allocated \acs{atb}, which has been tied to this particular client who sent the \acs{aths} request.
                It should be used in subsequent \acs{httpa}/2 requests to ensure those requests can be efficiently dispatched into \acsp{tservice}.
                Given that the \acs{httpa}/2 request dispatcher may not be trustworthy, and won't be capable to check its integrity of it.
                As a result, it cannot guarantee that those requests could be delivered into their matched \acsp{atb}. To remedy this problem, the dispatcher should be capable to identify invalid \acs{atb} ID as possible, and the receiving \acs{tservice} should validate it right after integrity check (see 4.2 in Figure.\ref{fig:secretprov}, 5.2 in Figure.\ref{fig:trusttx}).\\ 
                Note that the max-age directive set here indicates how long this \acs{atb} could be kept alive on the server side.
            \end{enumerate}
        \end{enumerate}
    \item \acs{atq} Exchange\label{itm:ra} \\
        In \acs{httpa}/2, a successful \acs{ra}~\cite{ra_review_2021} increases client's confidence by assuring the targeting services running inside a curated and trustworthy \acs{atb}.
        The client can also determine the level of trust in the security assurances provided by \acsp{tservice} through \acs{atb}. \\
        The \acs{ra} is mainly aimed to provision secrets to a \acs{tee}, In this solution, we leverage this mechanism to set it as the root trust of the \acs{httpa}/2 transactions instead of certificate-based trust, e.g., \acs{tls}.
        To facilitate it, we integrate the \acs{ra} with the key exchange mechanism above to perform a handshake, which passes the assurance to derived ephemeral key materials (see 3.9 in Figure.\ref{fig:raalloc}).
        Those keys can be in turn used to wrap secrets and sensitive data designated by the client or \acs{tservice} in either direction.\\
        During \acs{ra} process, the \acs{atq} (see section~\ref{attestquote}) plays a key role to attest \acs{tservice}. 
        It provides evidence (quote) to prove authenticity of the relevant \acs{tservice} (see 3.8 in Figure.\ref{fig:raalloc}).
        The client can just rely on it to decide whether the \acs{tservice} is a trustworthy peer or not~\cite{draftrats}.\\
        To appraise \acs{atq}, we need a trusted authority to be the verifier to perform the process of \acs{atq} verification, and reports issues on this \acs{atq}, e.g.,\acs{tcb} issues (see 3.12 in Figure.\ref{fig:raalloc}).
        The result of verification produced by the verifier should be further assessed by the client according to its pre-configured policy rules and applied security contexts.\\
        Importantly, the \acs{tservice} should ensure the integrity and authenticity of all \acsp{ahl} of \acs{atr} and its response through the \acs{qudd} of \acs{atq}, and vice versa in case of \acs{mhttpa}/2.\\
        The following \acsp{ahf} should be supported by \acs{httpa}/2 protocol for \acs{ra}.
        
        \begin{enumerate}
            \item \acsp{ahf} in request message (or \acs{atr}):
            \begin{enumerate}
                \item Attest-Quotes\\
                It can only appear in \acs{mhttpa}/2 mode to indicate a set of \acsp{atq} generated from the \acsp{tclient} for targeting \acs{tservice} to verify.
                (see 3.2, 3.3, 3.4, 3.5 in Figure.\ref{fig:raalloc}). 
                These quotes should be used to ensure \acs{intauth} of the \acsp{ahl} of this \acs{atr} through their \acs{qudd}.\\
                Note that the max-age directive indicates when these quotes are outdated and its cached verification results should be cleared up from \acs{atb} to avoid broken assurance.
                In addition, all client-side quotes must be verified by server-side verifier and validated by \acsp{tservice} before a \acs{atb} ID can be issued.
            \end{enumerate}
            
            \item \acsp{ahf} in response message
            \begin{enumerate}
                \item Attest-Quotes\\
                It is mandatory for a \acs{atb} to present its \acsp{atq} to the client for client-side verification.
                The \acs{intauth} of both \acsp{ahl} of the \acs{atr} and its response should be ensured by its \acsp{qudd} to protect the transaction completely.\\
                The client must verify the \acs{atq} to authenticate its identities of remote \acs{atb} (see 3.8, 3.10, 3.11, 3.12 in Figure.\ref{fig:raalloc}).
                The client should not trust anything received from \acs{tservice} before \acsp{atq} is successfully verified and evaluated.
                Whether the integrity of \acsp{ahl} is held should be determined by client-side security policies.
                Note that the \acs{tservice} quotes can be selectively encrypted in its parts through \acs{trc} to hide its identity information. 
            \end{enumerate}
        \end{enumerate}
    There are several remaining \acsp{ahf}, which are important to this transaction as they provide other necessary information and useful security properties:
    \begin{enumerate}
        \item \acsp{ahf} in request message
        \begin{enumerate}
            \item Attest-Versions\\
            The client presents an ordered list of supported versions of \acs{httpa} to negotiate with its targeting \acs{tservice}.
            
            \item Attest-Date\\
            It is the \ac{utc} when client initiates a \acs{aths}.
            
            \item Attest-Signatures\\
            It contains a set of signatures, which are used to ensure \acs{intauth} of \acsp{ahl} in this \acs{atr} through client-side signing key (see section~\ref{attestticket}).
            
            \item Attest-Transport\\
            As described in section~\ref{trustls}, the client HELLO message should be put in here.
            With this, the \acs{tservice} can enforce a trustworthy and secure connection at L5, which is a bit similar to what \ac{hsts} does~\cite{rfc6797}.
        \end{enumerate}
        
        \item \acsp{ahf} in response message
        \begin{enumerate}
            \item Attest-Version\\
            It shows client which version of \acs{httpa} is selected by \acs{tservice} to support.
            
            \item Attest-Transport\\
            Similarly, the \acs{tservice} returns its HELLO message to the client for a secure transport layer handshake.
            
            \item Attest-Expires\\
            It indicates when the allocated \acs{atb} will go expire and its related resources will get released. It provides another layer of security to reduce the chance of this \acs{atb} being attacked.
            
            \item Attest-Secrets\\
            It is an ordered list of \acs{atb}-wide secrets, which are provisioned by \acs{tservice} if client expects them. 
            This way can save a \acs{rtt} of \ac{atsp} (see section~\ref{ssec:secprov}) in case of \acs{tservice} won't demand secrets from client immediately.
            
            \item Attest-Cargo\\
            The usage of this field is described in section~\ref{trustcargo}. Note that ``Attest-Cargo'' is a \acs{ahf} while \acs{trc} is the corresponding content which plays an important role on sensitive data encryption and authentication.
        \end{enumerate}
    \end{enumerate}

    Apart from those tasks above, this \acs{atr} can act as a GET request, but it cannot be trusted due to incomplete key exchange at this moment, which means it cannot contains any sensitive data, but its response can be trusted as the key exchange process completed at \acs{tservice}-side, and before it gets returned. Therefore, the \acs{tservice}-side sensitive data can be safely transmitted back to the client through the \acs{trc}.

\end{enumerate}
\subsection{Attest Secret Provisioning (AtSP) Phase}\label{ssec:secprov}

\begin{figure*}[htp]
    \centering
    \includegraphics[width=0.8\textwidth]{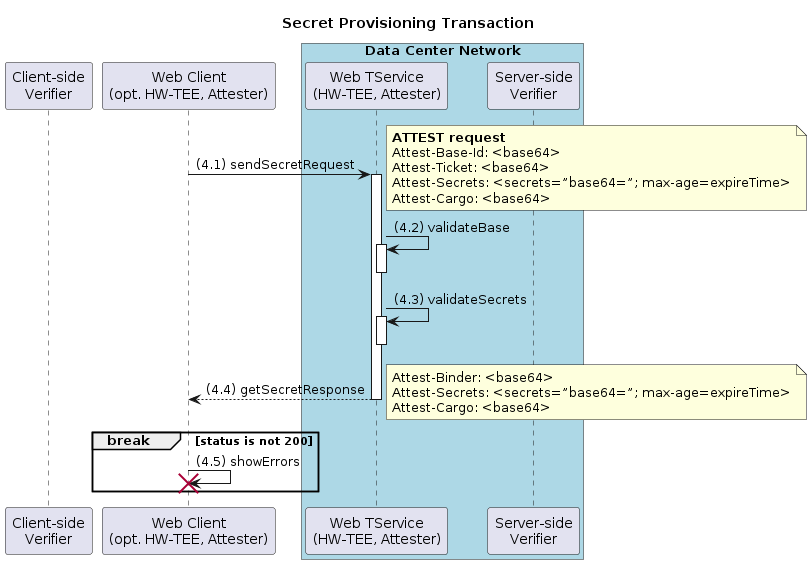}
    \caption{Attest secret provisioning (\ac{atsp}) transaction}
    \label{fig:secretprov}
\end{figure*}

As mentioned in section~\ref{itm:ra}, the main purpose of \ac{atsp} is to securely deliver secrets to a trustworthy \acs{atb}, which has been verified by a server-side verifier.
The \acs{atr} of \acs{atsp} is intended to be used for this purpose. 
To be precise, it is for \acs{atb}-wide and client-wide secret provisioning.
On the contrary, the request-wide or response-wide secrets should be carried by the \acsp{trc} (see section~\ref{trustcargo}) of \acs{httpa}/2 transactions. 
In addition, the failure of \acs{atsp} will causes \acs{atb} termination immediately.

As shown in Figure.\ref{fig:secretprov}, the \acs{atsp} transaction can be used to provision secrets in two directions since the \acs{atb} and its key materials already got derived through \acs{aths} (see section~\ref{sec:raalloc}) on both sides; thus, the \acsp{ahl} can be fully protected during this phase. 
Moreover, \acs{atr} of \acs{atsp} can be issued by the client any number of times at anytime after \acs{aths}.\\
These \acsp{ahl} described in following:

\begin{enumerate}
    \item \acsp{ahf} in request message (or \acs{atr})
    \begin{enumerate}
        \item Attest-Base-ID\\
        This identifier is used to specify which \acs{atb} is targeted to handle this \acs{atr} of \acs{atsp}. 
        With this ID, the \acs{tservice} can firstly validate it against its serving list to make sure correctly handling of this request (see 4.2 in Figure.\ref{fig:secretprov}).
        However, the \acs{tservice} should quietly ignore it if the ID is not valid for its residing \acs{atb} as the receiving \acs{tservice} should not expose any information for an adversary to exploit.
        
        \item Attest-Ticket\\
        The usage of this field is explained in section~\ref{attestticket}. 
        The value of this field must be unique to prevent a replay attack.
        Also, it ensures the \ac{intauth} of the \acsp{ahl} in this request.
        
        \item Attest-Secrets\\
        It contains an ordered list of secrets, which is wrapped up by means of \acs{ae} as a standard way for strong protection. 
        Moreover, each secret should be able to be referred to by the client later using the index. For example, specifying a provisioned secret that is used to decrypt embedded sensitive data.
        Again, the receiving \acs{atb} should be terminated if any of these provisioned secrets cannot be validated or accepted by the \acs{atb} (see 4.3 in Figure.\ref{fig:secretprov}).
        
        \item Attest-Cargo\\
        This field is optional, it can be used to carry any sensitive information, which is meaningful to \acs{tservice} (see section~\ref{trustcargo}).
        Note that this paper is not intended to define the structure of its content, which could be addressed in another one.

    \end{enumerate}
    \item \acsp{ahf} in response message:
    \begin{enumerate}
        \item Attest-Binder\\
        It is used to make sure the request to response is binding together to identify this transaction uniquely (see section~\ref{attestbinder}).
        
        \item Attest-Secrets\\
        In this \acs{hf}, these contained wrapped secrets will be provisioned back to the client. 
        As noted earlier, this can be merged into the response \acsp{ahl} in \acs{atr} of \acs{aths} (see section~\ref{sec:raalloc}).
        
        \item Attest-Cargo\\
        Similarly, it can be used to carry sensitive information/data back to the client (see section~\ref{trustcargo}).
    \end{enumerate}
\end{enumerate}

\subsection{Trusted Communication Phase}\label{ssec:trusttx}

\begin{figure*}[htp]
    \centering
    \includegraphics[width=0.8\textwidth]{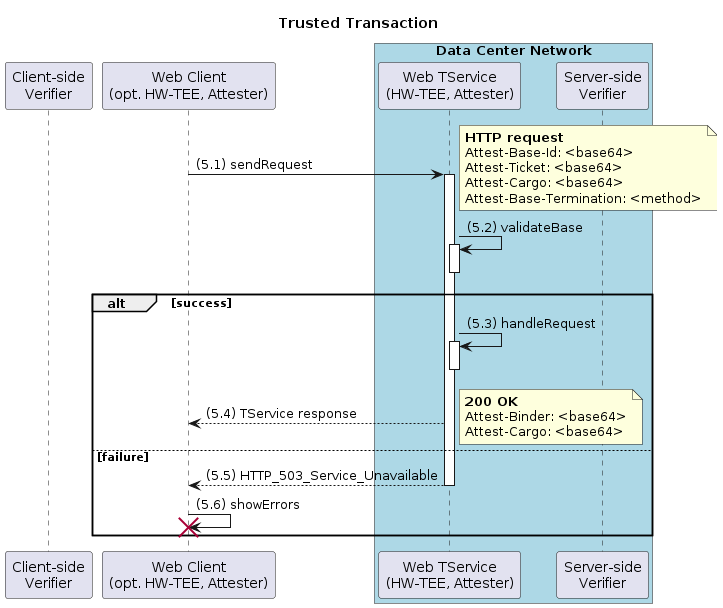}
    \caption{Trusted transaction}
    \label{fig:trusttx}
\end{figure*}

When \acs{atb} is allocated for the client, it can subsequently issue \acs{trr} (see section~\ref{trustreq}) to do the real work. 
Basically, the \acs{trr} is an ordinary \acs{http} request with some extra \acsp{ahl}, which are described in detail as follows:

\begin{enumerate}
    \item \acsp{ahf} in request message:
    \begin{enumerate}
        \item Attest-Base-ID\\
        It specifies which \acs{atb} to handle this request, and should be validated by targeting \acs{tservice} (see 5.2 in Figure.\ref{fig:trusttx}) before processing this request (see 5.3 in Figure.\ref{fig:trusttx}).
        
        \item Attest-Ticket\\
        This field has been explained above (see section~\ref{attestticket}), which is intended to authenticate this request, and prevent other \acsp{ahl} from being tampered with or being replayed.
        
        \item Attest-Cargo\\
        As noted earlier, this field is optional, and the client can use it to transfer arbitrary sensitive information to \acs{tservice} (see section~\ref{trustcargo}).
        
        \item Attest-Base-Termination\\
        We can include this \acs{ahf} if it is the last \acs{trr} towards the \acs{atb}.
        It is recommended way to terminate a \acs{atb} actively.\\
        The termination method can be one of the following options:
        \begin{description}
            \item[cleanup] \hfill \\
            This means that the terminated \acs{atb} can be reused by other clients. 
            
            \item[destroy] \hfill \\
            Specify this method, if the \acs{atb} should not be reused or shared by any other clients.
            
            \item[keep] \hfill \\
            This allows \acs{atb} to be shared with other clients. 
            Be careful, this method is less safe as the residual data could be exploited and leaked to the next client if any.
        \end{description}
    \end{enumerate}
    
    \item \acsp{ahf} in response message:
    \begin{enumerate}
        \item Attest-Binder\\
        As explained earlier, the \acs{httpa}/2 uses it to ensure the \acs{intauth} of both request and response together (see section~\ref{attestbinder}).
        
        \item Attest-Cargo\\
        As noted earlier, the \acs{tservice} can leverage this mechanism to transfer arbitrary sensitive information back to its client (see section~\ref{trustcargo}).
    \end{enumerate}
\end{enumerate}

\subsection{Protocol Flow}\label{ssec:protoflow}

\begin{figure}[htp]
    \centering
    \includegraphics[width=0.8\linewidth]{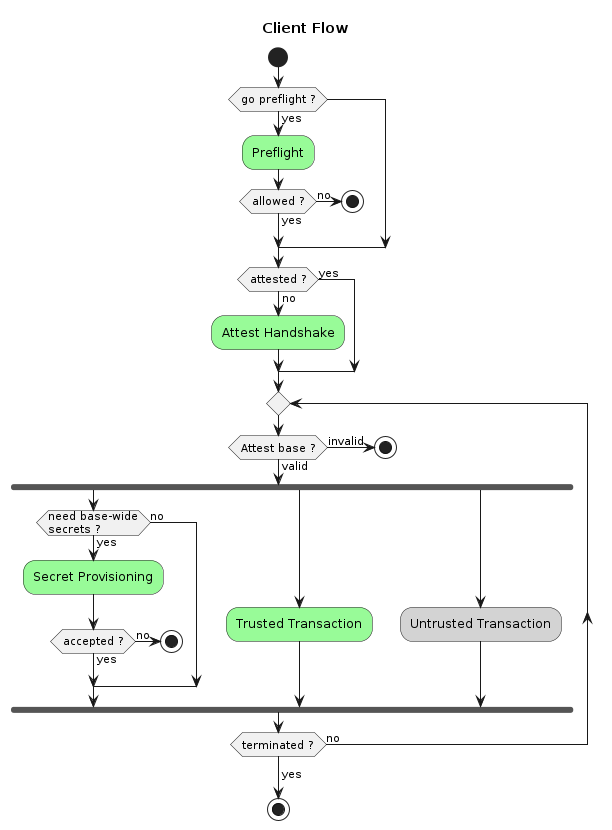}
    \caption{\acs{httpa} transaction workflow from the client view.}
    \label{fig:pflows}
\end{figure}

As shown in Figure.\ref{fig:pflows}, we illustrate those transactions from client perspective, including preflight, \acs{aths}, \acs{atsp}, and trusted request in a workflow diagram.
In the design of \acs{httpa}/2, only the phase of \acs{aths} is required, which not only largely simplifies the interaction between the client and the \acs{tservice} but also improves the \ac{ux}.

\begin{figure*}[htp]
    \centering
    \includegraphics[width=0.8\linewidth]{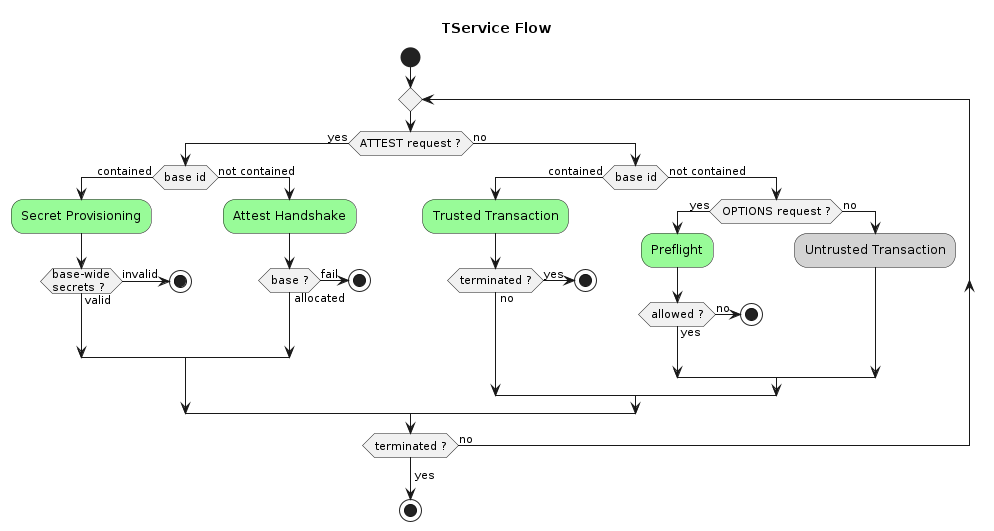}
    \caption{\acs{httpa} transaction workflow from the \acs{tservice} view}
    \label{fig:tsflows}
\end{figure*}
Figure.\ref{fig:tsflows} shows the workflow, which can help understand how those transactions are distinguished in \acs{tservice}.

%% file: tex/security.tex
In this section, we discuss security properties and the potential vulnerabilities, as is necessary for understanding \acs{httpa}/2. 

\subsection{Layer 7 End-to-End Protection}
In cloud computing scenarios, intermediary nodes, such as L7 load balancer or reverse proxy, are used commonly to improve the network performance to deliver the best web experience.
However, the great web experience does not come for free. 
The secure communication based on \acs{tls} only protects transmitted data hop-by-hop at layer 5 (L5).
The intermediary nodes may need TLS termination to inspect HTTP messages in plain text for better network performance.
As a consequence, the intermediary nodes can read and modify any HTTP information at L7.
Although \acs{tls} itself is not the problem, it cannot protect sensitive information above L5 where most Web services are located. 
That is the gap between L5 and L7 that causes the underlying vulnerability.
Therefore, the trust model including intermediary nodes, which are over L5, is problematic~\cite{rfc8613}, because, in reality, intermediary nodes are not necessarily trustworthy. 
Intermediary nodes may leak the privacy and manipulate the header lines of \acs{http} message.
Even in the case, where intermediaries are fully trusted, an attacker may exploit the vulnerability of the hop-by-hop architecture and lead to data breaches. 
\acs{httpa}/2 helps protect \acsp{ahl} and the sensitive information of HTTP message end-to-end at L7.
As long as the protection does not encrypt the necessary information against proxy operations~\cite{rfc8613}, \acs{httpa}/2 can provide guarantees that the protected message can survive across middleboxes to reach the endpoint.
Especially, \acs{httpa}/2 provides an encryption mechanism at the level of HTTP message, where only the selected information, some header lines or payloads, is encrypted rather than the entire message.
Thus, the parts of \acs{httpa} information without protection may be exploited to spoof or manipulate.
If we want to protect every bit of \acs{httpa} message hop-by-hop, \acs{tls} is highly recommended in combine with \acs{httpa}/2 for use.

In the implementation, the \acsp{tservice} can make a privacy policy to determine to what degree the \acs{httpa} message is protected to the L7 endpoint without \acs{tls} for better network performance. 
If the message is highly sensitive entirely, \acs{tls} can come to help in addition, but only up to the security of the L5 hop point. 
\subsection{Replay Protection}
A replay attack should be considered in terms of design and implementation.
To mitigate replay attacks, most AEAD algorithms require a unique nonce for each message.
In \acs{atr}, random numbers are used. 
In \acs{trr}, a sequential nonce is used on either endpoint accordingly. 
Assuming strictly increasing numbers in sequence, the replay attack can be easily detected if any received number is duplicated or no larger than the previously received number.
For reliable transport, the policy can be made to accept only \acs{trr} with a nonce that is equal to the previous number plus one. 

\subsection{Downgrade Protection}
The cryptographic parameters of configuration should be the same for both parties as if there is no presence of an attacker between them.
We should always negotiate the preferred common parameters with the peer.
If the negotiated parameters of configuration are different for both parties, it could make peers use a weaker cryptographic mode than the one they should use, thus leading to a potential downgrade attack~\cite{ bhargavan2016downgrade}. 
In \acs{httpa}/2, \acs{tservice} uses \acs{atq} to authenticate its identity and the integrity of the \acs{aths} to the client.
In \acs{mhttpa}/2, the client uses \acs{atq} carried by \acs{atr} for proving its own authenticity and the message integrity.
Thus, the communication traffic of the handshake across intermediaries cannot be compromised by attackers. 

\subsection{Privacy Considerations}
Privacy threats are considerably reduced by means of \acs{httpa}/2 across intermediary nodes.
End-to-end access restriction of integrity and encryption on the \acs{httpa}/2 \acsp{ahl} and payloads, which are not used to block proxy operations, aids in mitigating attacks to the communication between the client and the \acs{tservice}. 
On the other hand, the unprotected part of \acs{http} headers and payloads, which is also intended to be, may reveal information related to the sensitive and protected parts.
Then privacy may be leaked.
For example, the \acs{http} message fields visible to on-path entities are only used for the purpose of transporting the message to the endpoint, whereas the \acsp{ahl} and its binding payloads are encrypted or signed.
It is possible for attackers to exploit the visible parts of HTTP messages to infer the encrypted information if the privacy-preserving policy is not well set up.
Unprotected error messages can reveal information about the security state in the communication between the endpoints.
Unprotected signaling messages can reveal information about reliable transport.

The length of \acs{httpa}/2 message fields can reveal information about the
message.
\acs{tservice} may use a padding scheme to protect against traffic analysis.
After all, \acs{httpa}/2 provides a new dimension for applications to further protect privacy.

\subsection{Roots of Trust (RoT)}
Many security mechanisms are currently rooted in software; however, we have to trust underlying components, including software, firmware, and hardware.
A vulnerability of the components could be easily exploited to compromise the security mechanisms when the \acs{rot} is broken. 
One way to reduce that risk of vulnerability is to choose highly reliable \acs{rot}.
\acs{rot} consists of trusted hardware, firmware, and software components that perform specific, critical security functions~\cite{rootoftrust}. 
\acs{rot} is supposed to be trusted and more secure, so it is usually used to provide strong assurances for the desired security properties. 
In \acs{httpa}/2, the inherent \acs{rot} is the \acs{atb} or \acsp{tee}, which provide a firm foundation to build security and trust.
With \acs{atb} being used in \acs{httpa}/2, we believe that the risks to security and privacy can be greatly reduced.

%% file: tex/conclusion.tex
In this paper, we propose the \acs{httpa}/2 protocol, a major revision of \acs{httpa}/1.
\acs{httpa}/2 is a layer 7 protocol that builds trusted end-to-end communication between \ac{http} endpoints.
An integral part of \acs{httpa}/2 is based on confidential computing, e.g., \acs{tee}, which is used to build verifiable trust between endpoints with remote attestation.
Communication between trusted endpoints is better protected across intermediary nodes which may not have \acs{tls} protection.
This protection helps prevent \acs{httpa}/2 metadata and the selected \ac{http} data from being compromised by internal attackers, even with \ac{tls} termination.\\

In addition to security advantage, the \acs{httpa}/2 also illustrates the performance advantages over \acs{httpa}/1, as it is not mandatory to enforce \acs{tls}; hence the overheads of \acs{tls} can be saved. 
Furthermore, \acs{httpa}/2 provides flexibility for a service provider to decide which part of the \ac{http} message is required to be protected.
This feature can potentially be leveraged by \acsp{csp} to optimize their networking configuration and service deployment to improve the throughput and response time.
With those improvements, the energy of electricity can be saved as well.\\

%% file: tex/futurework.tex
To further realize \acs{httpa}/2 in the real world, we will be focused on \ac{poc} to demonstrate its validness and soundness.
We will apply the PoC codes of \acs{httpa}/2 to various use cases in practice.
Lastly, we plan to release a reference implementation towards generalization to open source. 

In the future, we expect emerging private AI applications to leverage \acs{httpa}/2 to deliver its end-to-end trusted service for processing sensitive data and protecting model IP. 
\acs{httpa}/2 will enable \ac{taas} for more trustworthy Internet.
With \acs{httpa}/2 and \ac{taas}, Internet users will have freedom of choice to trust, the right to verify assurances, and the right to know the verified details.
Users are able to make their decision out of free will based on the faithful results that they consider and choose to believe.
It becomes possible that we can build genuine trust between two endpoints.
Thus, we believe that \acs{httpa}/2 will accelerate the transformation process towards trustworthy Internet for a better digital world.